\documentstyle [aps]{revtex}
\begin{document}
\twocolumn

\title{ADDENDUM TO "NONLINEAR QUANTUM EVOLUTION\\ WITH MAXIMAL ENTROPY PRODUCTION"}
\author{S. Gheorghiu-Svirschevski\footnotemark[1]\footnotetext{e-mail: hnmg@soa.com}}
\address{1087 Beacon St., Suite 301, Newton, MA 02459}
\date{\today}
\maketitle

\begin{abstract}
The author calls attention to previous work with related results, which has escaped scrutiny before the publication of the article "Nonlinear quantum evolution with maximal entropy production", Phys.Rev.\textbf{A63}, 022105 (2001).
\end{abstract}

In a recently published paper \onlinecite{1}, I propose a nonlinear extension of the nonrelativistic Liouville-von Neumann equation, which incorporates both the unitary propagation of pure quantum states and the second principle of thermodynamics. Shortly after this work was approved for publication, it was brought to my attention \cite{2} that a closely related undertaking on the subject was reported about two decades ago by G.P.Beretta et al.\cite{3}-\cite{8}, and later also by Korsch et al. \cite{10,11}. The present Addendum is intended to acknowledge and correct this oversight, and to provide a few brief observations on the relation between these works and the results presented in \cite{1}.  

 In ref. \cite{3}, Beretta et al. proposed that the dynamical principle of quantum theory be replaced by a postulated nonlinear equation of motion which is, algebraically, a generalization of Eq.(24) in \cite{1} (see also Eq.(91)). General properties of this equation are also presented in an axiomatic framework, particularly with regard to the nature of nondissipative and equilibrium states (see also \cite{4}). Unfortunately, the applicability of the theory was hindered by a number of mathematical difficulties, e.g., by lack of a definitive proof of the positivity of the evolution. A  well-behaved example was given later for the two-level system in interaction with an external field \cite{5}. Such problems notwithstanding, it was argued \cite{6,7} that the proposed nonlinear evolution drives the density matrix along a direction of steepest entropy ascent under given constraints, and ref.\cite{7} provides a notable theorem on exact, generalized Onsager reciprocity, not restricted to the near-equilibrium regime. The stability of thermodynamic equilibrium states has also been discussed \cite{8}, with reference to (but not limited to the context of) the postulated nonlinear dynamics. More recently, Korsch et al. studied a family of closely related dissipative dynamics \cite{9}, and also derived explicit solutions for a driven harmonic oscillator by Lie-algebra techniques \cite{10}.

Beretta's confidence in the physicality of his construction seems to find vindication after all. In ref.\cite{1} the theory is formulated in terms of state operators $\gamma$ associated to the density matrix $\rho$ $[ \rho = \gamma \gamma^{+}]$ and the equation of motion is derived from a variational principle which observes the principles of quantum mechanics and the fundamental laws of thermodynamics. The existence and uniqueness of the solutions for $\rho$ follow from the equation of motion for $\gamma$, and the positivity of the evolution is guaranteed by construction. The reader can also find alternative proofs for the fundamental properties of the equation of motion, a derivation of a near-equilibrium limit, exact to first-order deviations of $\gamma$ from the equilibrium state, as well as a proof of the equivalence between symmetry covariance, conservation laws and associated commutation relations. 

The latter assertion implies that the generators for the nonlinear equation are determined by the dynamical symmetries of the system under consideration, and conveys a precise meaning for Beretta's "complete set of constants of motion". On the other hand, it must be noted that some constants of motion that Beretta includes in his set of "generators of motion", specifically observables associated with (nonrelativistic) superselection rules [e.g., the number of particles of a (physical) constituent], need not be accounted for separately. Indeed, let $\hat N$ be such an observable and $[\hat{N}, H] = 0$, with H the hamiltonian of the system. Any physically meaningful density matrix $\rho$ of an isolated system has non-vanishing matrix elements only between states corresponding to the same eigenvalue $\nu$ of $\hat N$, such that $\hat{N}\rho=\rho\hat{N}=\nu\rho$. But if $\rho$ evolves according to Eq.(24) in \cite{1}, it can be verified that $\hat{N}\dot{\rho}=\dot{\rho}\hat{N}=\nu\dot{\rho}$, which means that this property is preserved in time implicitly, and any additional constraint on the average of $\hat N$ is superfluous. 

Also, it appears that Beretta's approach to multi-component systems and separability \cite{11} falls outside the ansatz used in \cite{1}, and the associated mathematical difficulties remain to be solved. This is due to the occurrence of products of partial traces of the total density matrix $\rho$ over various subsystems, which prevent the reduction of the equation of motion for $\rho$ to an equation of motion for $\gamma$ . The point of view adopted in \cite{1} on separability  follows the conventional philosophy and does not distinguish between elementary and compounded systems, other than by their symmetry group. More precisely, the equation of motion is regarded as the expression of dynamical constraints imposed by symmetry principles, conservation of probability and the first and second principles of thermodynamics. Hence multi-component systems are on an equal footing with elementary systems, up to additional symmetries. In particular, noninteracting, mutually isolated systems display a supplementary time-translation symmetry, which restricts the accessible space of physical states to disentangled states, and consequently reduces (algebraically) both the conservation laws (including those generated by symmetry principles) and the second principle to corresponding statements for the isolated parts. The equation of motion that accounts for all the constraints in this case reduces to individual equations of motion for the parts, as expected for separable systems.

A last remark concerns Beretta's approach to the Onsager reciprocity relations \cite{7}. The reader may find it useful to note that Beretta's quorum of self-adjoint operators (a maximal set of linearly independent operators in the set of self-adjoint operators on the physical Hilbert space) is in fact a self-adjoint basis on the space of linear operators. Such bases always exist, and the set of self-adjoint (or better, hermitian) operators can be retrieved as the set of real linear combinations of the basis operators [for instance, a basis $\{|\alpha \rangle\}$ in the Hilbert space generates the basis $\{ |\alpha\beta) = | \alpha \rangle \langle \beta | \}$ for the operator space, which can be rearranged into the self-adjoint basis $\{ |\alpha\alpha), |\alpha\beta+), |\alpha\beta-) \}$, where  $| \alpha\beta+ )=( | \alpha \rangle \langle \beta | + | \beta \rangle \langle \alpha | ) / \sqrt{2}$  and $| \alpha\beta-)=i ( | \alpha \rangle \langle \beta | - | \beta \rangle \langle \alpha | ) / \sqrt{2} $]. Beretta's quorum corresponds to a generally non-orthogonal, non-normalizable, self-adjoint basis of observables, which can be obtained from a basis of the type shown here through a nonsingular linear transformation on the operator space. In this context, it can be easily verified that Beretta's approach to Onsager relations applies, in general, to evolution equations of the form 
\[ \dot{\rho} = \textbf{L}\log\rho + \frac{i}{\hbar}[\rho,\;H]  \]
with $\textbf{L}$ a tilde-symmetric superoperator \cite{1}, which may dependent nonlinearly on $\rho$, but acts linearly on $\log\rho$. Also, probability conservation [$Tr(\dot{\rho})=0$] demands $\textbf{L}^{+}I = 0$ for $I$ the identity on the wavefunction Hilbert space, while conservation of an observable $C$ [$Tr(C\dot{\rho})=0$] requires $\textbf{L}^{+}C = 0$. Assuming that the positivity of $\rho$ is also preserved, the entropy production will be positive if and only if $\textbf{L}$ is also hermitian [ $\textbf{L}^{+}=\textbf{L}$] and positive definite [$Tr(\alpha^{+}\textbf{L}\alpha) \geq 0$ for any $\alpha$ in the operator space].  In that case, for any self-adjoint (hermitian) operator basis $\{\chi_{\alpha} \}$, one can retrieve Onsager relations (see \cite{7}) for the Heisenberg transforms $\chi_{\alpha}^{H}=exp[(i/\hbar)Ht]\chi_{\alpha}exp[-(i/\hbar)Ht]$ and the corresponding coefficients  will be given by $L_{\alpha\beta}^{H}=Tr(\chi_{\alpha}^{H}\textbf{L}\chi_{\beta}^{H})$, with $L_{\alpha\beta}^{H}=L_{\beta\alpha}^{H}=(L_{\beta\alpha}^{H})^{*}$.

\vspace{0.5cm} 
Acknowledgement:  I remain indebted to Dr. Robert Englman of SOREQ, Israel for bringing Dr. Beretta's work to my attention. I also thank Dr. Beretta for providing a list of his publications.

\end{document}